\documentclass[journal=aamick,manuscript=article]{achemso}

\SectionNumbersOn

\usepackage{chemformula} 
\usepackage[T1]{fontenc} 
\usepackage{amssymb}
\usepackage{times}

\author{Dhiraj Nandyala}

\author{Zhen Wang}

\author{David Hwang}

\author{Thomas Cubaud}

\author{Carlos E. Colosqui}
\affiliation{Department of Mechanical Engineering, Stony Brook University, New York, NY 11794}
\alsoaffiliation[Stony Brook University]{Department of Applied Mathematics and Statistics, Stony Brook University, New York, NY 11794}

\email{carlos.colosqui@stonybrook.edu}

\title[Title]
{A capillary diode for potential application in water-oil separation}

\keywords{Capillary imbibition, selective adsorption, liquid-infused surfaces, water-oil separation}

\begin{document}

\begin{abstract}
A capillary device is designed and fabricated in glass to work as a fluidic diode with vanishingly small hydrodynamic conductance for imbibition of water within a finite range of immersion depths.
This is attained through patterning a section of predefined length on the device surfaces using a single-step laser-based ablation process and without resorting to chemical treatment of the hydrophilic glass substrate.
While the studied device works as a fluidic diode for water, it can behave as a conventional capillary slit for the imbibition of oils (e.g., alkanes, silicone oils) with low surface tension.    
A prototype device with simple geometric design is demonstrated for selective adsorption and separation of water and oil in vertical imbibition experiments at controlled immersion depths.
Efficient devices for passive separation of water and oil can be designed based on the demonstrated physical mechanism and the analytical model proposed in this work.
\end{abstract}


\section{Introduction}
%
%
Well-established thermodynamic arguments and experimental analysis indicate that the wetting behavior of different liquids can be controlled by designing the micro- or nanoscale structure or roughness of a solid surface.\cite{huh1977effects,oliver1980experimental,quere2008wetting,colosqui2017,yang2018underwater}
The micro- and nanoscale structure or roughness of a surface can significantly increase the interfacial area with direct contact with a wetting fluid and effectively increase the surface energy per projected unit area, as described by the Wenzel equation.\cite{wenzel1936} 
Depending on the affinity between the wetting fluids and the solid phase, the surface structure or roughness can entrap small domains of one fluid phase producing a heterogeneous surface with a dramatically different effective surface energy, as predicted by the Cassie-Baxter model.\cite{cassie1944} 
Furthermore, multiple local minima in the system energy and thus metastable equilibrium states can be induced by the micro/nanostructure of a surface, which has been found to substantially hinder or even prevent wetting processes such as spontaneous or forced spreading, capillary imbibition, or interfacial adsorption.\cite{kaz2012physical,colosqui2013,colosqui2015wetting,colosqui2016crossover,jose2018,colosqui2019}  
The dynamics of capillary imbibition can be reasonably well described by considering capillary pressure and viscous hydrodynamic resistance, as in the conventional Lucas-Washburn equation,\cite{washburn1921} for the case of macroscopically smooth and chemically homogeneous surfaces when the system is sufficiently far from equilibrium. 
However, capillary imbibition on micro/nanostructured surfaces presents nontrivial dynamic behaviors that are not accounted for by conventional models due to the presence metastable states and thermally activated processes induced by physical and/or chemical heterogeneities of wetted micro/nanopatterned surfaces. \cite{courbin2007imb,gruener2009spontaneous,xue2014switch,zheng2020super}
Engineering the micro/nanostructure of a surface can thus provide diverse pathways to control the static and dynamic wetting behavior of different liquids such as water and oil.

In the case of wetting by water in ambient air, proper combinations of surface chemistry and physical structure produce macroscopically observed superhydrophobic \cite{feng2002super,long2015superhydrophobic,tian2016moving,aktar2020} or superhydrophilic \cite{drelich2011hydrophilic,zhang2014fabrication,otitoju2017superhydrophilic} behaviors that are characterized by extremely high ($\gtrsim 150^\circ$) or low ($\lesssim 10^\circ$) apparent contact angles, respectively.   
The contact angle hysteresis can become negligibly small for certain combinations of surface micro/nanostructure and geometric configurations (e.g., sessile droplet, static meniscus), which leads to very small liquid adhesion forces and the associated self-cleaning and antifouling effects.\cite{long2015superhydrophobic,kobayashi2012wettability,hoshian2015robust} 
Analogous wetting phenomena can be observed for the case of oils wetting a solid, in which case micro/nanostructured surfaces can display superoleophobic or superoleophilic properties.\cite{kota2012hierarchically,zhang2014surface,yong2018review,baruah2019synthesis}  
However, the fabrication of surfaces with robust and stable omniphobic properties that prevent adhesion of different liquids and oils presents significant challenges that cannot be easily overcome due to surface contamination and chemical transformations over long times with associated undesirable effects.

Efforts to overcome these challenges led to the development of liquid-infused surfaces (LISs) with the micro- or nanoscale structure of the surface filled with liquid having high affinity to the surface material.\cite{wong2011bioinspired,lafuma2011slippery,solomon2014drag,sett2017lubricant}
Under proper conditions, the liquid filling the micro/nanostructure can form a stable thin film preventing direct contact between the wetting liquid and effectively providing a lubrication effect that significantly reduces contact angle hysteresis and adhesion forces.
The macroscopic wetting behavior of a LIS is thus effectively controlled by the affinity between the infused liquid and the wetting fluid phases, which can result in a large variety of wetting configurations.\cite{preston2017design,villegas2019liquid,peppou2020life}  
While the wetting properties of LISs have been widely studied for the case of open-surface configurations (e.g., droplet spreading on a single surface), a relatively small number of studies have focused on the case of full confinement (e.g., two-phase flow in microchannels) where the coupling between pressure, gravitational, and capillary forces can give rise to diverse and complex wetting dynamics.\cite{bico2002wetting,jacobi2015overflow,wexler2015shear,keiser2020universality}  

In this work, we develop and verify a theoretical model for imbibition considering metastable equilibrium states induced by micro/nanoscale surface features and demonstrate a relatively simple and low-cost fabrication technique without resorting to chemical treatment to induce long-lived metastable states at predicted locations that fully prevent or hinder the imbibition of specific liquid pairs (e.g., water-air). 
The employed fabrication technique is suitable for glass, which has the mechanical, chemical, and thermal stability required for numerous engineering applications. 
We thus design, fabricate, and characterize a glass slit microchannel with a section of prescribed length occupied by a LIS and a second section with a plain (non-patterned) surface.
Under proper conditions, a large surface energy barrier is induced at the junction between the channel section with the LIS and the plain conventional surface, which effectively results in a finite threshold pressure for driving net liquid flow trough the channel. 
The energy barrier magnitude and minimum pressure head required to sustain fluid flow can be controlled by the geometric design of the slit channel and the microstructure of the LIS, and critically depends on the surface energy of the wetting fluids.
The designed slit microchannel can be employed as a microfluidic check-valve or ``capillary diode'' with different critical pressures heads for conducting water or oils through capillary action and pressure-driven flows.
To demonstrate the potential application of the concept for the passive separation of water and oil, we characterize experimentally the vertical capillary imbibition behavior of a simple prototype device in the case of water or hexadecane in ambient air, and the more complex case of oil droplets and films on a water phase. 
%
%
Our experimental observations can be accounted for by a simple analytical model that outlines the basic principles for the design of so-called capillary diodes as potential devices for selective adsorption and separation of immiscible liquids.

\section{Theoretical background \label{sec:theory}}
\begin{figure}[h!]
\centering
\includegraphics[width=0.55\textwidth]{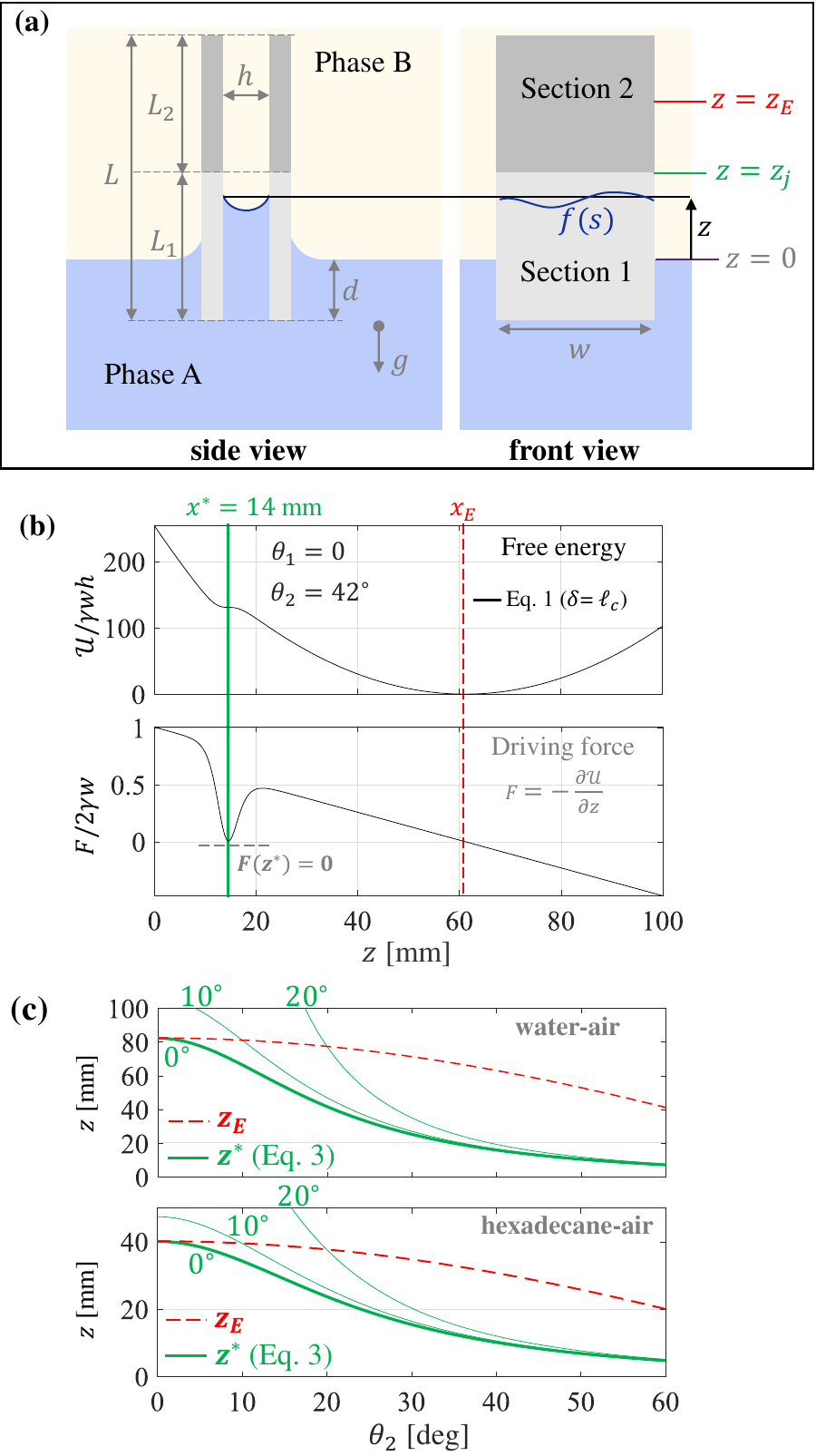}
\caption{Studied capillary device and analytical model predictions. 
(a) Vertical imbibition of a capillary slit with two sections having Young contact angles $\theta_1$ and $\theta_2$ prescribed by their effective surface energies.
(b) Dimensionless system free energy and driving force predicted by Eq.~\ref{eq:energy} for water-air ($\theta_1=0$ and $\theta_2=40^\circ$) and a slit channel ($w\gg h$) with the section junction at $z_j=z^*=14$ mm ($\delta=\ell_c$).
(c) Critical junction position $z^*$ for producing metastability predicted by Eq.~\ref{eq:xs} for $\theta_1=$ 0$^\circ$, 10$^\circ$, and 20$^\circ$ and $\theta_2=$ 0-60$^\circ$ in the case of water ($\rho=997$~kg/m$^3$, $\gamma=72$ mN/m) or hexadecane oil ($\rho=770$~kg/m$^3$, $\gamma=27$ mN/m) in ambient air.
}
\label{fig:1}
\end{figure}

The studied capillary device (Fig.~\ref{fig:1}a) consists of a 3D slit channel of length $L$ with a rectangular cross section of width $w$ and narrow gap height $h\ll w$, that is filled with two immiscible fluids A and B having an interfacial tension $\gamma$.
The slit channel has two sections having different effective surface energies $\gamma_{Ai}$ and $\gamma_{Bi}$ ($i=1,2$) when wetted by fluid A and B, respectively.
Hence, the surface energy difference $\Delta\gamma_i=\gamma_{Ai}-\gamma_{Bi}$ determines the Young contact angle $\cos\theta_{i}=-\Delta\gamma_i/\gamma$ for section 1 and 2. 
The two sections of the slit have lengths $L_1$ and $L_2$, respectively, and the junction between the two sections is located at a distance $z_j=L_1-d$ from the flat interface of the bath at $z=0$ when the leading edge of the slit is immersed at a depth $d$ (Fig.~\ref{fig:1}a).

This analysis focuses on the case of vertical capillary imbibition against the action of gravity (Fig.~\ref{fig:1}a) when the studied device is immersed into a large bath of fluid A and $\theta_1\le\theta_2<90^\circ$.
As the fluid phase A displaces phase B, the average position of the three-phase contact line $z=(1/{\cal S})\times\int_0^S f(s) ds$ is determined by averaging the local contact line position $f(s)$ (Fig.~\ref{fig:1}a) along the full contact line perimeter $s=0,{\cal S}=2(w+h)$.  
We will consider that the closed system comprising the capillary slit and the fluid phases A and B is in thermal and chemical equilibrium. 
Furthermore the fluids A and B are considered to be incompressible with a constant mass density $\rho_A$ and $\rho_B$, respectively.
The system free energy parameterized by the average contact line position $z$ can be thus expressed as
\begin{equation}
\mathcal{U}(z)=\frac{1}{2} \Delta\rho g w h z^2+\gamma wh \frac{(\pi-2\theta)}{2\cos\theta}+2\Delta\gamma(z) (w+h) z
\label{eq:energy}
\end{equation}
where $\Delta\rho=\rho_A-\rho_B$ is the mass density contrast, $g$ is the gravitational acceleration, $\theta$ is the apparent contact angle at the three-phase contact line, and 
\begin{equation}
\Delta\gamma(z)=
-\gamma\cos\theta_1
+\frac{\gamma}{2}(\cos\theta_1-\cos\theta_2)
\left[1+\tanh\left(\frac{z-z_j}{\delta}\right)\right]
\label{eq:dgamma}
\end{equation}
is the surface energy difference, which is modeled as having a gradual spatial variation centered at the junction coordinate $z_j$ and over a finite distance 
$\delta= (1/{\cal S}) \int_0^{\cal S} (f(s)-z)^2 ds$.
The characteristic amplitude of the spatial variation $|f(s)-z|$ of the local contact line position is determined by a balance between surface and gravitational energy and thus we have $\delta=\ell_c g(\theta)$, 
where $\ell_c=\sqrt{\gamma/\Delta\rho g}$ is the capillary length and $g(\theta)\sim{\cal O}(1)$ is a shape factor close to unity for $w\gg \ell_c$ and $\theta \ll 1$.

Under the modeled conditions, the effective driving force for the displacement of the average contact line position is $F(z)=-\partial{\cal U}/\partial z$. 
The global minimum of the free energy ${\cal U}$ in Eq.~\ref{eq:energy}, and thus the stable equilibrium condition, is located at $z_E=(2\gamma\cos\theta_2/\Delta\rho g)\times(1/h+1/w)$ provided that $z_j$ is sufficiently smaller than $z_E$.
Moreover, Eqs.~\ref{eq:energy}--\ref{eq:dgamma} predict a local energy minimum at the critical value 
\begin{equation}
z^*\simeq
(\cos\theta_1+\cos\theta_2)\times
{\left[\frac{\Delta\rho g}{\gamma(1/h+1/w)} +\frac{(\cos\theta_1-\cos\theta_2)}{\delta} \right]}^{-1}.
\label{eq:xs}
\end{equation}
As shown in Fig.~\ref{fig:1}b for the particular case of $\theta_1=0$ and $\theta_2=40^\circ$, when the position of the junction $z_j\ge z^*$ is equal or larger than the critical value estimated by Eq.~\ref{eq:xs} the capillary imbibition of liquid will stop at a metastable equilibrium position for which $F(z^*)=0$ and there will be no further conduction of fluid into section 2. 
Such a metastable equilibrium at $z^*<z_E$ preventing further capillary conduction will be observed for very long observation times, provided that the energy barrier to cross the junction $\Delta{\cal U}\simeq 2\gamma |\cos\theta_2-\cos\theta_1| (w+h) z_j\gg k_B T$ is orders of magnitude larger than the thermal energy $k_B T$ and the assumption of chemical and thermal equilibrium is valid (e.g., when evaporation and surface condensation is not significant).   
It is worth noticing that for a given liquid pair and surface micro/nanopattern with specific geometric features determining the contact angle contrast ($\theta_1$-$\theta_2\neq 0)$, the energy barrier $\Delta{\cal U}$ can be readily adjusted as it varies linearly with the slit width $w$ and gap height $h$, as well as the junction position $z_j$.

The critical junction position for metastable equilibrium, as predicted by Eq.~\ref{eq:xs}, is reported in Fig.~\ref{fig:1}c for the case of water (or hexadecane oil) in a microscale capillary slit with a gap height $h=180$~$\mu$m and width $w\gg h$ that has a superhydrophilic and hydrophilic (or superoleophilic and oleophilic section) surface section for which $\theta_1\simeq$ 0, 10, or 20$^\circ$ and $\theta_2=$~20-60$^\circ$, respectively, when $\delta=\ell_c$ in Eq.~\ref{eq:dgamma} is determined by the capillary length for each liquid pair.      
The stable equilibrium $z_E$ and metastable equilibrium position $z^*$ in Fig.~\ref{fig:1}c are reported for a range of different contact angle values $\theta_2$ that can be adopted on the plain surface due to the phenomenon of static contact angle hysteresis.\cite{dussan1979spreading,petrov1991comparison,eral2013contact}
According to analytical predictions (Fig.~\ref{fig:1}c), placing the junction position at 20 mm above the flat interface will ensure that capillary imbibition of water is stopped at the junction for a wide range of contact angles $\theta_1\lesssim 20^\circ$ and $\theta_2\ge 35^\circ$, which are within the range of values commonly observed for water on micropatterned or conventional glass surfaces.
For both micropatterned or conventional glass surfaces, equilibrium contact angles for hexadecane or other oils are expected to be significantly lower ($\theta_1\simeq\theta_2\le 20^\circ$) and the capillary conduction of oil will be able continue above the junction until reaching the stable equilibrium state at $z_E>z^*$ or fill the entire slit when $L \le z_E$ (Fig.~\ref{fig:1}c).

Hence, analytical predictions from Eq.~\ref{eq:xs} for the particular conditions reported in Fig.~\ref{fig:1}c indicate that a simple prototype device with $z_j \simeq 20$~mm can prevent the vertical capillary imbibition of water into the second section where $z>z_j$, while oil can readily flow over the junction and fill the second section of the slit.  
Both water or oil can flow over the junction if its height above the flat bath interface is reduced to $z_j=10$ mm, which can be readily accomplished by immersing the edge of the device at $d=10$ mm inside the bath (Fig.~\ref{fig:1}a). 
This analytical finding suggests a possible strategy for separation of water from mixtures of water and less dense oils dispersed on its surface, by immersing the analyzed capillary device at controlled depths for which different fluid phases are able to flow continuously through the slit channel.
In particular, a device with a first superhydrophilic section filled with water can be potentially employed as a fluidic diode preventing spontaneous or forced imbibition by any liquid phase below a threshold pressure head $\Delta p^*=\Delta\rho g (z_E-z^*)>0$ across the slit channel. 
The validity of the presented analytical predictions will be assessed through the fabrication of the studied device and its experimental analysis reported in the following sections.

\section{Experimental methods}
We employ a facile fabrication method involving laser ablative patterning of glass to produce the studied fluidic capillary device consisting of a microscale slit with two sections having different effective surface energies, and thus different equilibrium contact angles $\theta_{1}$ and $\theta_{2}\gtrsim\theta_{1}$, for different liquid pairs (e.g., water-air, oil-air, and water-oil).
The equilibrium contact angles and their hysteresis are experimentally determined by optical goniometry and Wilhelmy plate measurements. 
The prototype device is designed according to the model presented in Sec.~\ref{sec:theory} to work as capillary diode or check valve with a critical pressure head $\Delta p^*$, attained at a critical immersion depth $d^*$ that is significantly different for the cases of water and oil in ambient air, which thus enables the selective capillary conduction of these liquids.   

\subsection{Fabrication}
\begin{figure*}[]
\centering
  \includegraphics[width=\textwidth]{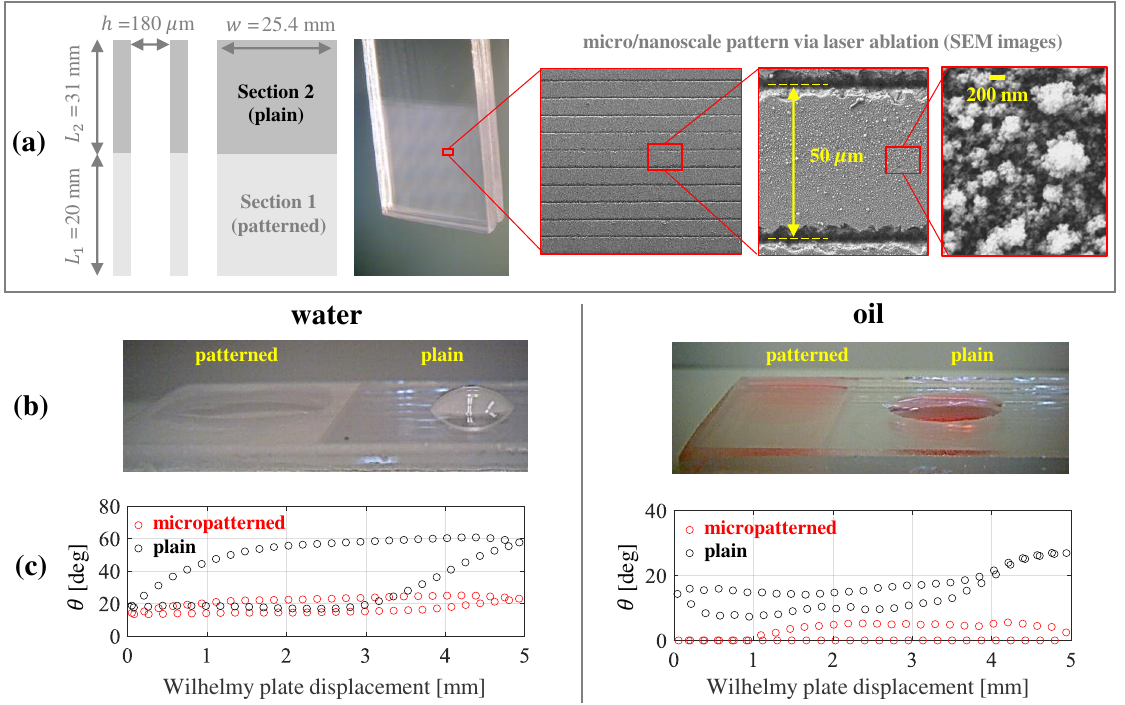}
\caption{Fabricated capillary device and surface wetting behavior. 
(a) Glass slit (gap height $h=180~\mu$m) having a 20-mm-long section with micro/nanoscale surface patterning consisting of microgrooves uniformly separated at distance $s=50~\mu$m and nanoscale surface features produced by laser ablation (SEM images). 
(b) Sessile droplet and wetting film formation for water and hexadecane oil observed on the plain (non-patterned) and micropatterned surface sections for water and oil.
(c) Contact angle (advancing/receding) hysteresis for water and hexadecane oil on the plain and micropatterned surface sections obtained using the Wilhelmy plate method during immersion/withdrawal.}
\label{fig:2}
\end{figure*}

The fabricated device consists of a rectangular slit channel with a nominal gap height $h=180\pm 5$~$\mu$m, width $w=25.4$~mm, and length $L=50.8$~mm (Fig.~\ref{fig:2}a) that is made of two borosilicate glass slides separated by spacers.
The fabrication begins with the laser ablation process \cite{choi2002femtosecond,hwang2006efficiency,hwang2009three} on both sides of the two borosilicate glass slides (i.e., inner and outer channel surfaces) to produce a micro/nanopatterned surface section of length $L_1=20$ mm as reported in Fig.~\ref{fig:2}a. 
A picosecond laser (RG10-H, Photonic Industries International Inc.) of 532 nm in wavelength, $~$12 ps in full width half maximum temporal pulse width, 100 kHz in the pulse repetition frequency and up to 7W in average laser power, is used for patterning the glass surfaces. 
The laser beam was focused on the glass surface with an infinity-corrected objective lens of 0.14 in numerical aperture (Plan APO 5X, Mitutoyo) to the laser spot diameter of $~$10 $\mu$m by 1/e definition. 
The glass surfaces are scanned with a fixed speed of 20 mm/s using precision XY motion stages (LS103H-100-XY, Aerotech Inc.) and a fixed laser power of 400 mW.
The laser power is precisely adjusted by an external attenuator set composed of a half waveplate and a polarizing beam splitter. 
A zoom lens (12X zoom lens system, Navitar Inc.) and a charge-coupled device (CCD) camera (XC-75, SONY) were implemented at the back of the objective lens to precisely locate laser focus on the sample surface and in-situ monitor the laser machining process. 
The surface pattern consists of parallel microgrooves of depth $d_g\simeq 7$~$\mu$m and width $w_g\simeq 10$~$\mu$m that cover the entire slit width with a uniform spacing $s=50$~$\mu$m.
Nanoscale features of small linear dimensions ($\sim$10 to 100 nm) that fully cover the patterned glass surface are deposited during the laser ablation process.
The patterned slides are bonded together using spacers and sealed with epoxy to produce a rectangular slit channel open on both ends.

\subsection{Surface wetting characterization \label{sec:wetting}}

The contact angles for the patterned (section 1) and plain (section 2) surfaces of the fabricated slit were determined using the sessile drop and the Wilhelmy plate methods for the case of water and hexadecane oil in ambient air, as reported in Figs.~\ref{fig:2}b-c. 
For the reported measurements the glass slides were cleaned with DI water and anionic detergent (Alconox) in an ultrasonic bath sonicator, followed by isopropyl alcohol and DI water rinse.
Deionized (DI) water and hexadecane stained with Oil Red O for visualization were employed for the measurements in Figs.~\ref{fig:2}b-c. 
The surface tension measured using the Du Nouy ring method in ambient air before and after the experiments was $\gamma=72\pm 5\%$ mN/m for DI water and $\gamma=28\pm 5\%$ mN/m for hexadecane; the addition of Oil Red O did not produce noticeable changes in the surface tension measured for the hexadecane oil.
For the contact angle measurements with the sessile drop method, a controlled droplet volume $V_d=10$~$\mu$L was deposited using a pipette on the patterned and plain (non-patterned) surface and the equilibrium contact angle was determined from digital images using the software ImageJ\cite{abramoff2004image}, with 8 to 10 repeated measurements to determine average values.
For the case of water (Figs.~\ref{fig:2}b) on the patterned surface (section 1) thin wetting films were formed with low apparent contact angles that we were not able to determine accurately with the employed optical method, while sessile droplets formed with a contact angle $\theta_2=41^\circ\pm 9\%$ measured on the plain glass surface (section 2).
For the case of hexadecane oil (Figs.~\ref{fig:2}b), while again the formation of thin films with extremely low apparent contact angle on the patterned surface precluded an accurate determination of the apparent contact angle, a small but finite contact angle $\theta_2\simeq 15^\circ\pm 10\%$ was determined on the plain surface. 
Contact angle values $\theta_1\simeq 21.4^\circ$ for water and $\theta_1=0$ for oil on the patterned surface can be analytically estimated with the Wenzel equation $\cos\theta_1=r\cos\theta_2$ using the roughness factor $r=(s+2d_g+w_g)/(s+w_g)=1.23$ determined by the ratio of the contact area to the projected area for the fabricated microgroove pattern, with the spacing $s=$~50 $\mu$m, and the nominal microgroove depth $d_g$~7 $\mu$m and width $w_g=$10 $\mu$m. 

Further experimental characterization of the contact angle and its static hysteresis on each section of the fabricated slit was performed with the Wilhelmy plate method (Fig.~\ref{fig:2}c) at a low plate speed $V=$~5 mm/min that corresponds to extremely small capillary numbers $Ca \sim 10^{-6}$ and ``quasi-static'' conditions with negligible hydrodynamic effects.
Force measurements reported in this work were obtained with a Sigma 700 force tensiometer (Biolin scientific) using as Wilhelmy plates the partially patterned borosilicate slides employed to build the capillary slits. 
Wilhelmy plate measurements at sufficiently slow plate speeds enable the determination of the static advancing and receding contact angles $\theta_A$ and $\theta_R$, respectively, with the advancing contact angle $\theta_A>\theta_R$ observed during the immersion of the plate. 
Our measurements (Fig.~\ref{fig:2}c) report a negligible contact angle hysteresis $\theta_A-\theta_R\lesssim 5^\circ$ on the patterned surface section for both cases of wetting by water and oil in ambient air.
As expected, a significant contact angle hysteresis $\theta_A-\theta_R\simeq 35^\circ$ is observed for the case of water wetting the non-patterned surface section.
For the case of water in air, the micro/nanopattern produces not only a significant reduction of the equilibrium contact angle $\theta_1$ (cf. Fig.~\ref{fig:2}b) but also the suppression of contact angle hysteresis on the first section of the fabricated capillary slit device (cf. Fig.~\ref{fig:2}c).
The observed wetting behavior indicates that the micro/nanopatterned surface is infused by the wetting liquid (i.e., water or oil), which remains within the slit after multiple successive immersion and withdrawal cycles performed during the Wilhelmy plate measurements.
Moreover, analytical predictions for the presence of a metastable state (Fig.~\ref{fig:1}c) preventing imbibition of water remain valid when considering the observed hysteresis range and the measured contact angles $20^\circ\lesssim \theta_2 \lesssim 60^\circ$ on the plain surface and $\theta_2\simeq 20 \pm 2^\circ$ on the patterned surface section. 

\section{Results and discussion \label{sec:results}}
\subsection{Vertical imbibition}
Based on the equilibrium contact angles determined for water in Sec.~\ref{sec:wetting}, the studied capillary diode device is designed and fabricated with a patterned surface section of length $L_1=20$ mm in order to prevent the imbibition of water in ambient air for a significant range of immersion depths $0\le d \le d^*$, where the critical immersion depth $d^*=L_1-z^*\simeq$ 5-8 mm is estimated by Eq.~\ref{eq:xs} with $\delta/\ell_c\simeq$~0.8-1.1. 
The estimated critical immersion depth corresponds to a threshold pressure head $\Delta p^*=\Delta\rho g (z_E-z^*)\simeq$ 0.46-0.5 kPa for vertical imbibition of water. 
In the case of hexadecane oil for which $\theta_1\simeq\theta_2$, Eq.~\ref{eq:xs} gives $d^*\simeq 0$ and the fabricated device is expected to behave as a conventional capillary slit at any immersion depth.

\begin{figure*}[]
\centering
  \includegraphics[width=0.95\textwidth]{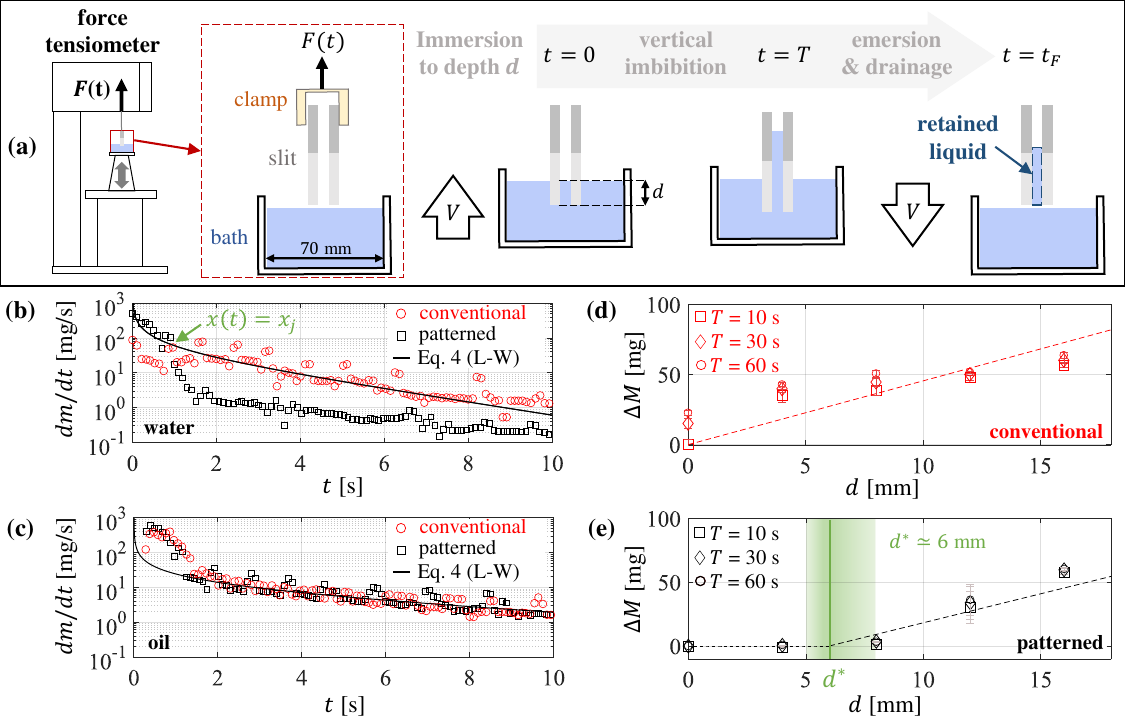}
\caption{Vertical imbibition and adsorption of water and oil. 
(a) Experimental setup for vertical imbibition experiments with immersion at controlled depths $d$ during a time $T$ followed by removal from the liquid bath. The immersion/withdrawal speed is $V=5$~mm/min in all cases. 
(b)-(c): Instantaneous mass rate $dm/dt$ at $d=0$ for water and hexadecane oil in a conventional capillary slit and the fabricated micropatterned device.
(d)-(e): Retained water mass increase $\Delta M=m(t_F)-m_{ref}$ (see definition in the main text) in the conventional capillary slit and micropatterned device after immersion times $T=$~10, 30, and $60$ s at immersion depths $d=$ 0, 4, 8, 12, and 16 mm.
The critical immersion depth $d^*=6$~mm determined from experimental measurements is predicted by Eq.~\ref{eq:xs} for $\delta=\ell_c$.
The shaded region in panel (e) shows a range of analytical predictions for $d^*$ using $\delta/\ell_c\simeq$~0.8-1.1. 
}
\label{fig:3}
\end{figure*}

Vertical imbibition experiments with DI water or hexadecane oil in ambient air (Fig.~\ref{fig:3}) are performed by immersing at precisely controlled depths $d$ the fabricated capillary device with a micro/nanopatterned section and a conventional capillary slit with the same nominal dimensions.
The experimental setup employed is illustrated in Fig.~\ref{fig:3}a, the capillary slit is hooked to the force tensiometer with a motorized stage to lift and lower the liquid bath (positioning resolution 0.016 $\mu$m) that enables the automatic and precise detection of the flat bath interface ($z=0$) and control of the immersion depth $d$.
The tensiometer force is zeroed at the beginning of the experiment when the studied sample and clamp arrangement is outside the liquid bath (Fig.~\ref{fig:3}a). 
The liquid bath is then lifted slowly at a slow constant speed $V=$~5 mm/min until reaching the desired immersion depths $d=$~0 to 16 mm, after which the tensiometer force $F(t)$ is automatically recorded over time at sampling intervals $\Delta t\simeq$ 0.1 s. 
The slit remains statically immersed at a depth $d$ for a time interval $T$ after which it is removed outside the bath, completing a full immersion-emersion cycle at the final time $t_F$ (Fig.~\ref{fig:3}a).
The imbibed liquid mass $m(t)=F(t)/g$ and the retained liquid mass $m_F=F(t_F)/g$ after removal from the bath are thus determined from the measured tensiometer force, and thus includes a very small additional contribution from the liquid film wetting the outer surfaces of the slit.

As expected for the case of zero immersion depth $d=0$ (Figs.~\ref{fig:3}b-c) and after a short initial time, the imbibed mass rate $dm/dt$ for water ($\rho_A=997$ kg/m$^3$) or hexadecane oil ($\rho_A=770$ kg/m$^3$) follows the conventional Lucas-Washburn (L-W) prediction,\cite{washburn1921,zhmud2000,fries2008analytic}  
\begin{equation}
\frac{dm}{dt}=\rho_A wh\frac{dz}{dt}=-\frac{\rho_A h^2}{12\mu z}\frac{d{\cal U}}{dz},
\end{equation}
determined from the L-W equation $z-z_E\log(1-z/z_E )=(\Delta\rho g h^2/12 \mu) \times t$, which neglects inertial effects and considers that plane Poiseuille flow develops inside the capillary slit.
For the fabricated capillary device with the micropatterned surface section, the water imbibition rate decays sharply after the water-air meniscus reaches the section junction at $z(t)=z_j$, as shown in Fig.~\ref{fig:3}b. 
The observed decay in the imbibed water mass rate is attributed to pinning forces caused by the steep change of the effective surface energy $\Delta \gamma$ near the end of the micropatterned section, as analytically predicted by Eqs.~\ref{eq:energy}--\ref{eq:dgamma}. 
This pinning effect is not observed for the imbibition of oil (Fig.~\ref{fig:3}c), for which the estimated change in surface energies is less significant between the patterned and plain slit sections.

\begin{figure*}[t!]
\centering
\includegraphics[width=0.95\textwidth]{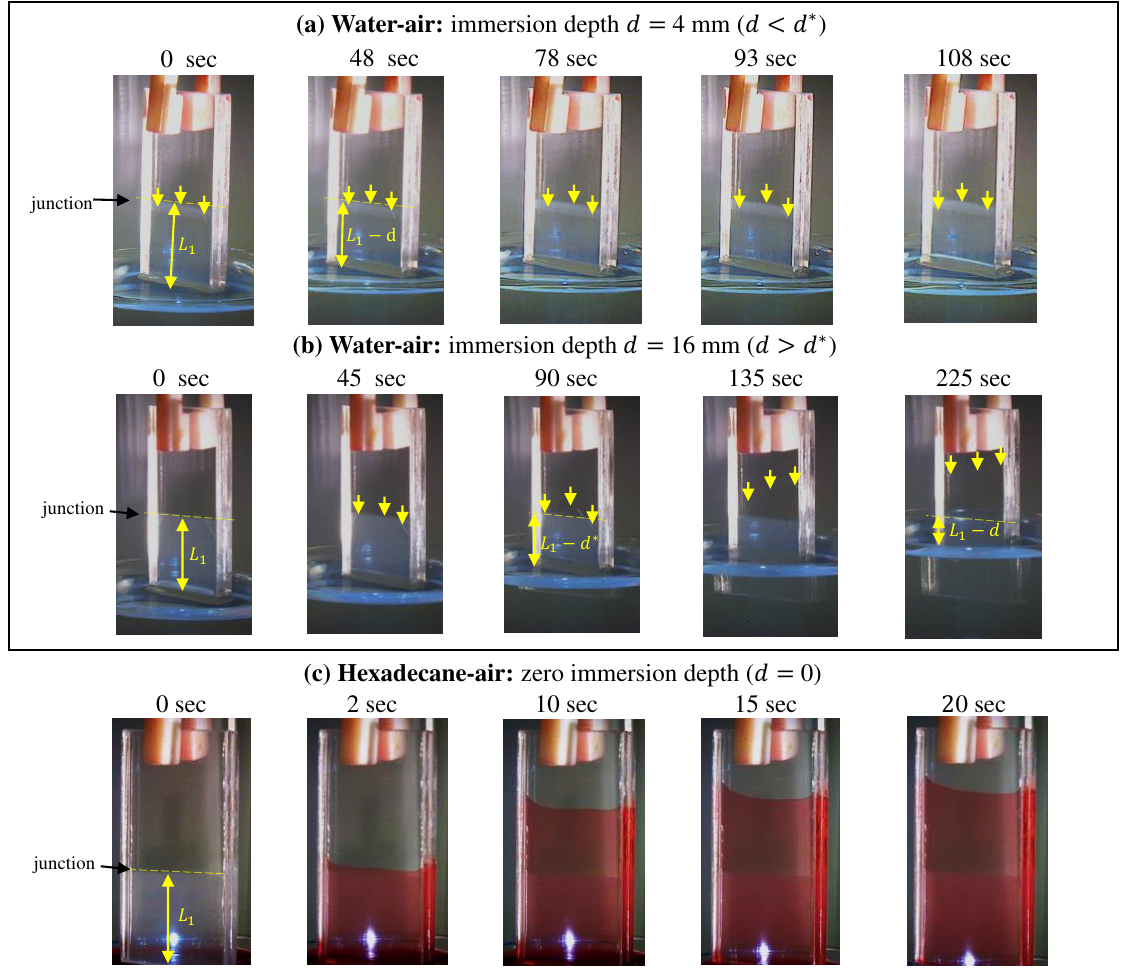}
\caption{
Vertical imbibition of the micropatterned capillary device at immersion depths above/below the critical immersion depth $d^*\simeq 6$ mm (plate speed $V=5$~mm/min).
Image sequences are snapshots at different time instances ($t\ge$~0) of Videos S1-S3 in the Supporting Information.
The position $z_j$ of the junction between the patterned and plain surface section is indicated by the dashed (yellow) lines.
Vertical (yellow) arrows are included to aid visualization of the water-air interface position.
(a) Water-air imbibition for $d<d^*$ [Video S1].
The water-air interface is not able to cross the section junction during an observation period of 1 min of static immersion at $d=4$~mm.
(b) Water-air imbibition for $d>d^*$ [Video S1].
The water-air interface crosses the section junction when $z_j\simeq L_1-d^*$ ($t=90$~s), before reaching the targeted immersion depth $d=16$~mm.
(c) Hexadecane-air system at zero immersion depth $d=0$ [Video S3].
The hexadecane oil is stained with Oil Red O for visualization.
The oil phase crosses the section junction at $z_j$ for any immersion depth 
$d\ge 0$.
\label{fig:4}} 
\end{figure*}

To quantitatively compare the water mass retained by a conventional capillary slit and the micropatterned device when immersed at depth $d=$~0, 4, 8, 12, and 16 mm over a finite time $T=$~10, 30, and 60 s, we report in Figs.~\ref{fig:3}d-e the mass increase 
$\Delta M(d,T)=m(t_F)-m_{ref}$, where the reference mass is $m_{ref}=M(0,10 s)$ for the conventional capillary slit and $m_{ref}=\rho_A w h L_1$ for the studied device, which corresponds to the water mass to fill the micropatterned section of the studied device. 
The experimental measurements reported in Figs.~\ref{fig:3}d-e correspond to averages over 3 to 5 realizations of immersion-emersion cycles under the same experimental condition, with error bars indicating 90\% confidence intervals.
For the case of a conventional capillary slit (Fig.~\ref{fig:3}d) and sufficiently long immersion times $T\ge 10$ s we observe an expected nearly linear increase of the retained mass $\Delta M \simeq \rho_A w h d$ that is caused by increasing the immersion depth. 
In contrast, the designed capillary device (Fig.~\ref{fig:3}d) fills up to the edge of the micropatterned section in less than 10 sec and prevents the further conduction of water above the section junction at $z_j=L_1-d$ until the immersion depth reaches a critical value $d^*\simeq$ 6 mm. 
The observed critical immersion depth for forward conduction of water in the micropatterned device falls within the analytical estimates for $d^*$ obtained via Eq.~\ref{eq:xs} (shaded region in Fig.~\ref{fig:3}e).
The retained mass increase $\Delta M$ and effective mass adsorption rate $\dot{M}=\Delta M/\Delta T$ can be accounted for by a model analogous a simple piecewise linear model for diode behavior\cite{agarwal2005foundations,leenaerts2013piecewise}, with a ``turn on'' immersion depth $d^*$ and effective forward ``conductance'' 
$\Delta \dot{M}/\Delta d= \rho_A w h/\Delta T$ for $d\ge d^*$ in the case of sufficiently long immersion times $T> 10$ s.

In addition to the quantitative analysis in Fig.~\ref{fig:3} direct visualization is reported in the image sequences in Fig.~\ref{fig:4} and Videos S1-S3 in the Supporting Information (SI). 
Video imaging confirms that the rising water-air meniscus remains static at the edge of the micropatterned slit section when the leading edge of the device is immersed at $d=4$ mm into the water bath (Fig.~\ref{fig:4}a and Video S1).
As expected, the water phase is able to cross the junction and flow into the second section when increasing the immersion depth to $d=16$ mm (Fig.~\ref{fig:4}b and Video S2 in SI), which is larger than the predicted critical immersion depth.
In contrast, for the vertical imbibition of hexadecane oil, the oil phase is always able to cross the junction at $z_j=L_1-d$ for $d\ge 0$ and flow into the second section, as reported in Fig.~\ref{fig:4}c (Video S3 in SI).  
Given that the diode-like behavior for the imbibition of water for different immersion depths does not occur for the imbibition of hexadecane oil, the micropatterned capillary device can be employed to selectively adsorb water and oil by controlling the immersion depth $d$. 
This suggests the possible application of the studied device for the separation of water and oil through properly devised strategies, as discussed in the next section.

\subsection{Water-oil separation}
\begin{figure*}[]
\centering
  \includegraphics[width=1\textwidth]{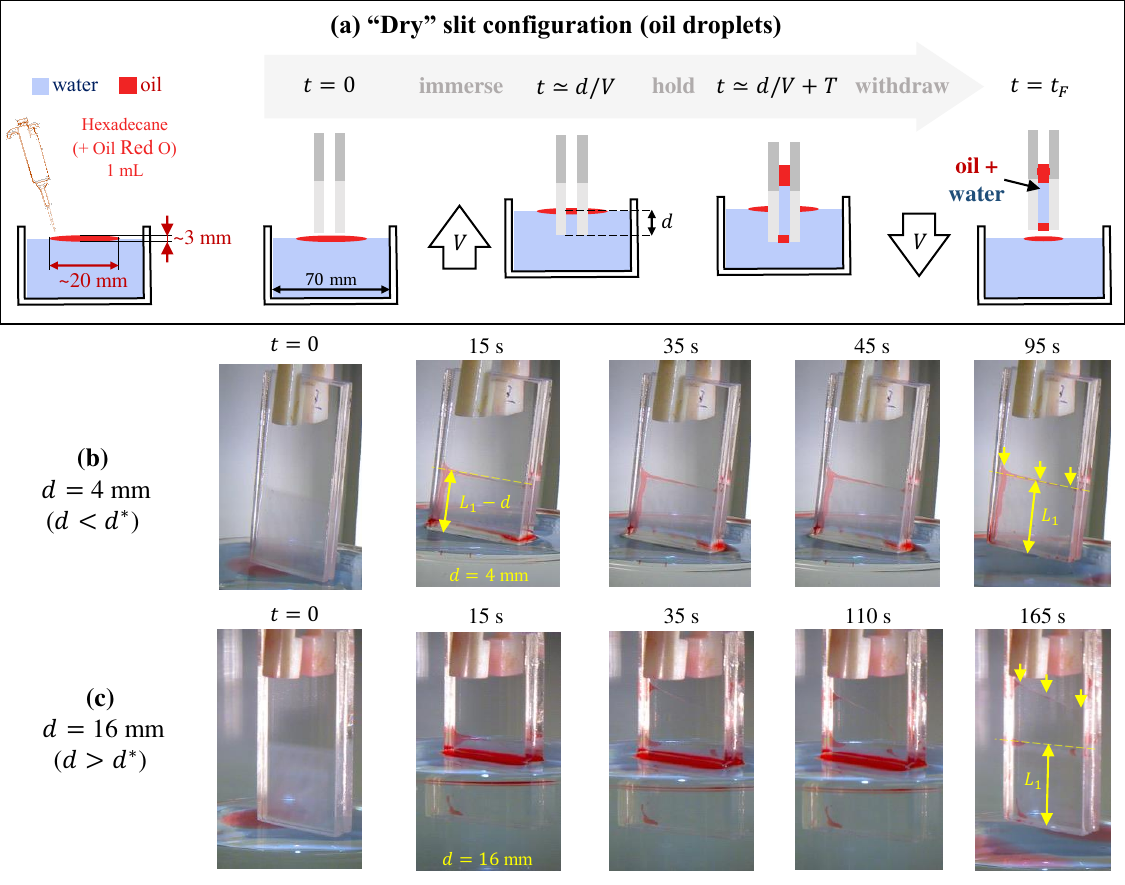}
\caption{Oil-water imbibition behavior in ``dry'' device configuration.
(a) Dry slit configuration where the micropatterned device does not contain liquid at the beginning of the immersion-emersion cycle at controlled depths $d$. 
The plate speed is $V=20$~mm/min in all cases.
An oil droplet of diameter $\simeq$20 mm and thickness $\sim$3 mm is deposited on the water-air interface by drop casting 1 mL of hexadecane oil (stained with Oil Red O for visualization).
(b) Immersion at depth $d<d^*$ smaller than critical ($d=4$ mm). 
The image sequence (Video S4 in SI) shows that both water and oil are adsorbed inside the slit within the patterned section.
(c) Immersion at depth $d>d^*$ larger than critical ($d=16$ mm)
The image sequence (Video S5 in SI) shows that both water and oil are adsorbed inside the slit above and below the junction between the patterned and plain section.
Yellow arrows are visual aids to indicate the position of the water-air interface.
.}
\label{fig:5}
\end{figure*}

\begin{figure*}[]
\centering
  \includegraphics[width=1\textwidth]{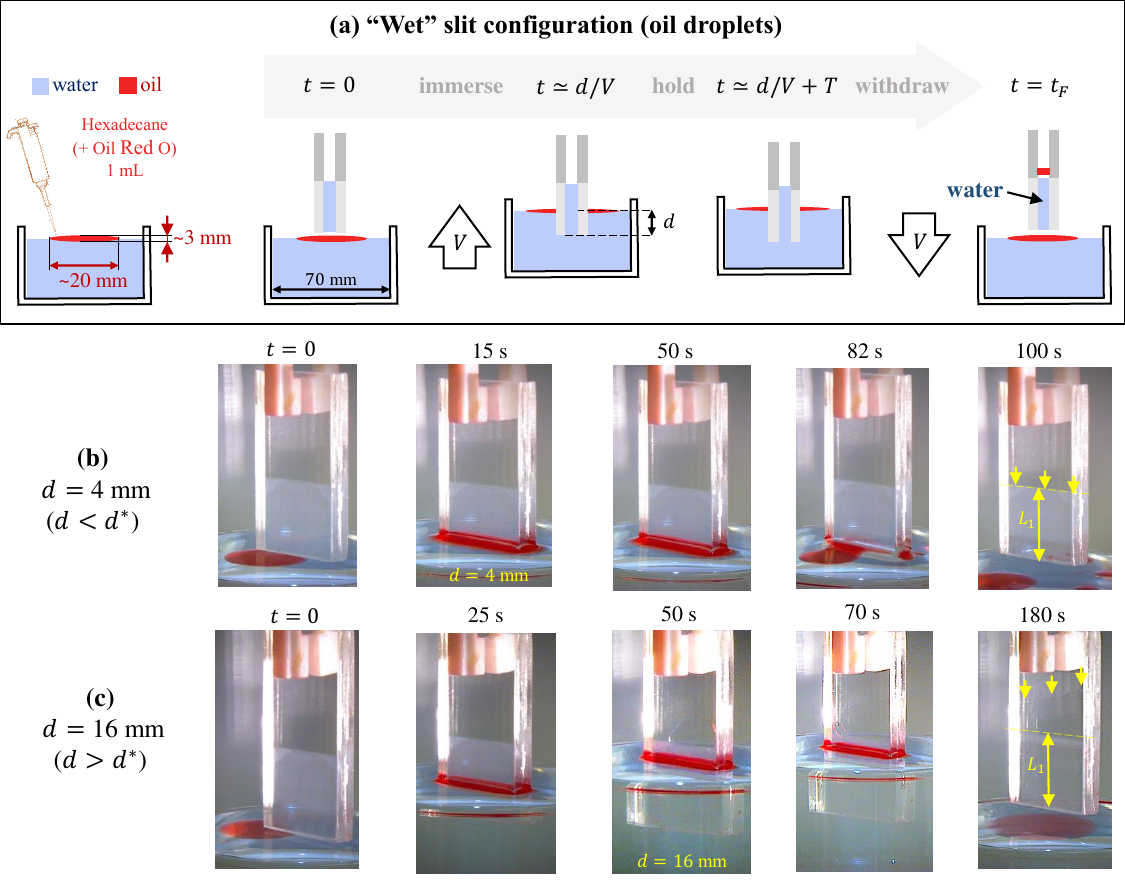}
\caption{Oil-water imbibition and separation in ``wet'' device configuration.
(a) Wet slit configuration where the micropatterned device contains water within the patterned section at the beginning of the immersion-emersion cycle at controlled depths $d$. 
The plate speed is $V=20$~mm/min in all cases.
An oil droplet of diameter $\sim$20 mm and thickness $\sim$3 mm is deposited on the water-air interface (hexadecane oil stained with Oil Red O for visualization).
(b) Immersion at depth $d<d^*$ smaller than critical ($d=4$ mm). 
The image sequence (Video S6 in SI) shows that only water is adsorbed inside the slit and within the patterned section only.
(c) Immersion at depth $d>d^*$ larger than critical ($d=16$ mm)
The image sequence (Video S7 in SI) shows that only water is adsorbed inside the slit both above and below the junction between the patterned and plain section.
Yellow arrows are visual aids to indicate the position of the water-air interface.}
\label{fig:6}
\end{figure*}

\begin{figure*}[]
\centering
  \includegraphics[width=1\textwidth]{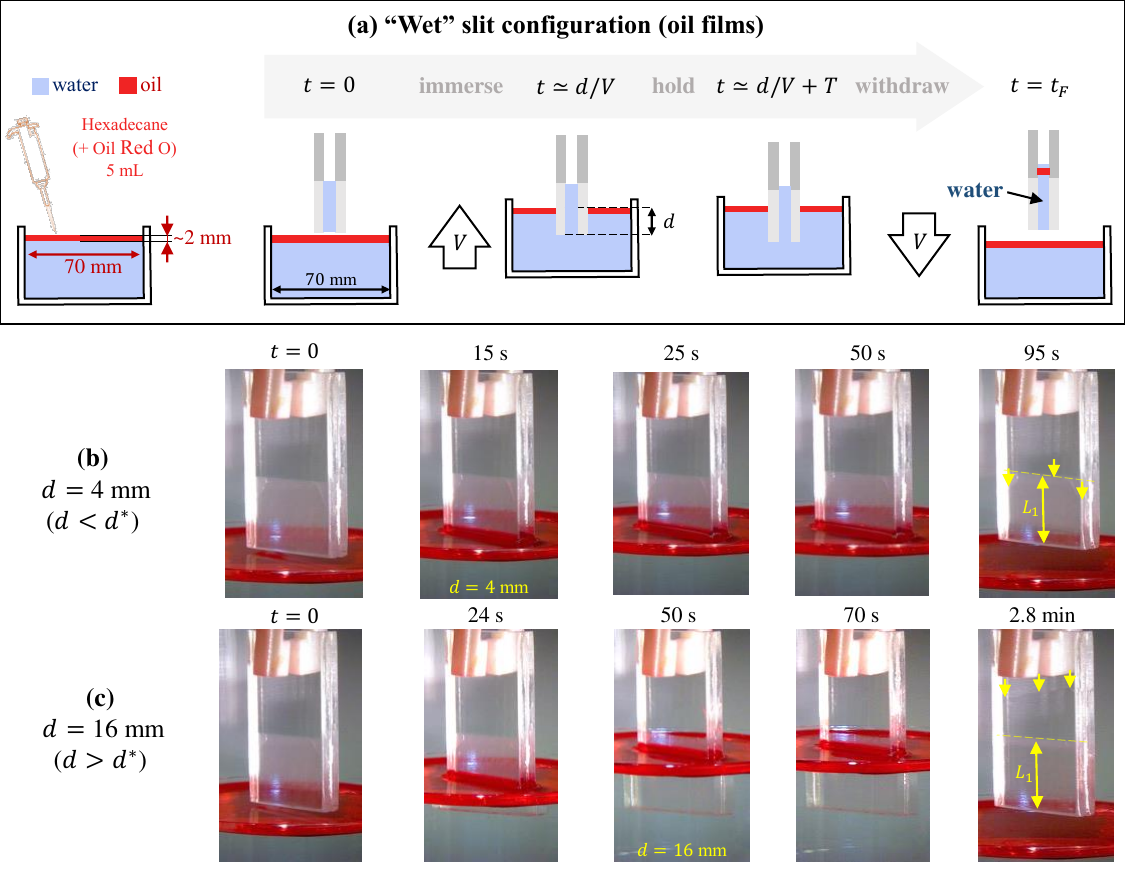}
\caption{Oil-water imbibition and separation in ``wet'' device configuration (oil films on water).
(a) Wet slit configuration where the micropatterned device contains water within the patterned section at the beginning of the immersion-emersion cycle at controlled depths $d$. 
The plate speed is $V=20$~mm/min in all cases.
An oil film of thickness $\sim$3 mm is deposited on the water-air interface (hexadecane oil stained with Oil Red O for visualization).
(b) Immersion at depths $d<d^*$ smaller than critical ($d=4$ mm). 
The image sequence (Video S8 in SI) shows that only water is adsorbed inside the slit and within the patterned section only.
(c) Immersion at depths $d>d^*$ smaller than critical ($d=16$ mm)
The image sequence (Video S9 in SI) shows that only water is adsorbed inside the slit above and below the junction between the patterned and plain section.
Yellow arrows are visual aids to indicate the position of the water-air interface.}
\label{fig:7}
\end{figure*}

In this section we assess the potential for water-oil separation of the studied micropatterned capillary device displaying diode-like behavior in the case of water imbibition (cf. Figs.~\ref{fig:3}--\ref{fig:4}).
For this purpose, vertical imbibition experiments are performed as illustrated in Figs.~\ref{fig:5}--\ref{fig:7} with the studied capillary diode device immersed in a water bath with different deposited volumes of hexadecane oil (1 to 5 mL), which produces oil droplets of diameter ($\sim$ 20 mm) comparable the slit width or continuous films with a characteristic thickness $\sim$ 2 to 3 mm.
In all studied cases, immersion speeds ($V$= 5 to 20 mm/min) lower than the capillary imbibition speeds (~1-10 mm/s) predicted by Eq. 4 for $z\le z_j$ are employed to ensure the validity of the theoretical analysis in Eqs. 1 to 4 and the observation of a metastable state preventing further imbibition at $z=z^*$ (Eq. 3). 
For higher immersion speeds approaching the capillary imbibition speeds, dynamic effects could force the liquid to flow over the position where metastable state is predicted by Eq. 3.

The studied device is first in a ``dry'' configuration (Fig.~\ref{fig:5}) where the patterned slit section is empty of any liquid, and a ``wet'' configuration (Figs.~\ref{fig:6}--~\ref{fig:7}) where water fills entirely the patterned section, which produces the reported diode-like behavior with a critical immersion depth $d^*\simeq 6$ mm for ``forward'' liquid conduction.  
The studied device in dry and wet configurations is immersed in the water bath with hexadecane oil (stained with Oil Red O for visualization) at a slow speed $V=20$~mm/min until reaching different immersion depths $d<d^*=4$ mm and $d>d^*=16$ mm above and below the critical immersion depth for water imbibition.

As reported in the image sequences in Fig.~\ref{fig:5}, the slit in the dry configuration adsorbs both water and oil droplets when the immersion depth is $d<d^*$ (Fig.~\ref{fig:5}b) or $d>d^*$ (Fig.~\ref{fig:5}c), for which both oil and water phases cross the patterned section junction. 
Hence, the dry-slit configuration leads to the combined imbibition of water and oil and does not produce the effective separation of the liquid phases.

On the other hand, the imbibition cases using the wet configuration for which the patterned surface section of the slit is infused with water prior to immersion result in the selective adsorption of water alone as shown in Figs.~\ref{fig:6}--\ref{fig:7}.
Notably, only water is adsorbed within the slit and the oil is not able to flow into the slit neither for the case of droplets smaller than the slit width (Fig.~\ref{fig:6}) or continuous oil films (Fig.~\ref{fig:7}) of thickness smaller than the critical immersion depth $d^*\simeq 6$ mm.
For immersion depths $d<d^*$ smaller than the critical value (see Fig.~\ref{fig:6}b\&\ref{fig:7}b), the water-air interface remains ``pinned'' at the junction between the patterned and plain section as previously observed for water-air systems, which prevents the imbibition of oil.
For sufficiently large immersion depths $d>d^*$, and once the oil phase is crossed, the water phase alone is able to flow over the junction and fill entirely the slit (see Fig.~\ref{fig:6}c\&\ref{fig:7}c).
Hence, when immersing the studied device pre-filled with water (i.e., ``wet'' configuration) the water phase can be effectively separated from the oil through selective and spontaneous capillary adsorption.  

\section{Conclusions}
In this work we have developed and validated a theoretical model for imbibition on micro/nanostructured surfaces and demonstrated a simple theoretical model and fabrication technique to design and build capillary devices with diode-like behavior that prevents the imbibition of certain liquid pairs for a finite range of immersion depths of the device. 
Notably, this was accomplished using a one-step laser ablation method producing simultaneously microscale patterning and formation of nanostructures on (hydrophilic) glass surfaces without any chemical treatment (e.g., hydrophobic silane-based coatings). 
The use of glass is highly desirable in diverse microfluidic applications due to its thermal, chemical, and mechanical stability, and optical transparency with low spectroscopic noise [see new references]. Moreover, the micro/nanostructures glass surfaces produced in this study are washable and reusable.

We have performed an analytical and experimental study of a prototype capillary device consisting of a microscale slit channel with two sections having different surface energies when wetted by different liquid pairs.
The contrast in surface energies, characterized by the apparent or effective contact angle was produced by the physical micro/nanopatterning, which when wetted by water or oil produces a LIS occupying a slit section of predetermined length and a junction with a plain (non-patterned) surface located at a specific position $z_j$ measured from the slit entrance.
The simple analytical model employed in this work can be used to design the slit channel dimensions and length of the section occupied by the LIS in order to induce a metastable equilibrium state for which the net force driving the conduction of different immiscible fluids becomes vanishingly small. 
Due to the metastable equilibrium at the junction between the patterned and non-patterned sections, the studied device can act in manner analogous to a capillary diode with asymmetric capillary conduction of liquid in the forward (imbibition) and backward (drainage) directions. 
This so-called capillary diode device permits the conduction of liquid water or oil in the ``forward'' direction ($z>0$) above very different critical immersion depths, which results in their selective adsorption under properly devised conditions. 

The properties of the studied capillary diode device can be exploited for the separation of water and oils and other immiscible fluid pairs having a significant contrast in their surface energies, which can be amplified by the physical micro/nanopattern and/or chemical treatment.
A strategy for passive separation of water and oil in vertical imbibition was demonstrated using a simple device produced by a facile fabrication procedure involving a single-step laser ablation for the physical micro/nanoscale patterning of transparent glass surfaces.
For the case of water in ambient air, the fabricated surface micropattern with uniformly spaced straight microgrooves and deposition of nanoscale surface features produced sufficient surface energy contrast to observe diode-like behavior in vertical capillary imbibition.
According to the analytical model employed in this work, it is feasible to attain diode-like behavior and enhanced conduction selectivity for different liquid pairs (e.g., water, fuels and oils) by employing alternative surface patterns with more complex geometries and nanoscale slit dimensions, and additional chemical treatment of the surface using hydrophobic and/or oleophobic coatings.

As observed for the studied vertical imbibition configuration, the studied capillary diode device can be employed to extract water from a mixture with a less dense oil through a passive process driven by capillary action.
Forced imbibition with controlled hydrostatic pressure differences could be employed to enhance the volumetric flow rate and/or possibly separate water from a mixture with a highly dense and viscous oil (e.g., heavy oils) provided that the micro/nanoscale pattern employed remains infused by water during the process, which is an open question requiring further experimental studies.
Devices for oil-water separation with enhanced throughput, determined by the volumetric rate of water separated from an oil-water mixture or emulsion, could consist of arrays with large number of the studied micro/nanopatterned capillary slits stacked together.
Furthermore, the separation efficiency could be significantly improved by reducing the gap height of the capillary slit channel, which prescribes the minimum diameter of oil droplets that can flow into the device when the micro/nanopatterned section is infused with water.
Future experimental analysis is required to further verify the analytical predictions proposed in this work and assess the performance of prototype devices designed for concrete technical applications by using proper performance metrics.

\begin{suppinfo}

The supporting information includes nine digital video sequences (S1-S9) with the full image sequences reported in Figs.~4-7:
\begin{itemize}
\item S1.mp4: Water-air imbibition with pre-filled micropatterned section immersed above critical depth ($d<d^*$)
\item S2.mp4: Water-air imbibition with pre-filled micropatterned section immersed below critical depth ($d>d^*$)
\item S3.mp4: Hexadecane oil-air imbibition with micropatterned slit at zero immersion depth $d=0$
\item S4.mp4: Water-oil (droplet) imbibition with initially dry micropatterned section immersed above critical depth ($d<d^*$)
\item S5.mp4: Water-oil (droplet) imbibition with initially dry micropatterned section immersed below critical depth ($d>d^*$)
\item S6.mp4: Water-oil (droplet) imbibition with pre-filled micropatterned section immersed above critical depth ($d<d^*$)
\item S7.mp4: Water-oil (droplet) imbibition with pre-filled micropatterned section immersed below critical depth ($d>d^*$)
\item S8.mp4: Water-oil (film) imbibition with pre-filled micropatterned section immersed above critical depth ($d<d^*$)
\item S9.mp4: Water-oil (film) imbibition with pre-filled micropatterned section immersed below critical depth ($d>d^*$)
\end{itemize}

\end{suppinfo}

\begin{acknowledgement}
This work was supported by the National Science Foundation through award CBET-1605809.
This work used resources of the Center for Functional Nanomaterials, which is a U.S. DOE Office of Science Facility, at Brookhaven National Laboratory under Contract No. DE-SC0012704. 
\end{acknowledgement}


\providecommand{\latin}[1]{#1}
\makeatletter
\providecommand{\doi}
  {\begingroup\let\do\@makeother\dospecials
  \catcode`\{=1 \catcode`\}=2 \doi@aux}
\providecommand{\doi@aux}[1]{\endgroup\texttt{#1}}
\makeatother
\providecommand*\mcitethebibliography{\thebibliography}
\csname @ifundefined\endcsname{endmcitethebibliography}
  {\let\endmcitethebibliography\endthebibliography}{}





\end{document}